\documentclass[%
 reprint,
superscriptaddress,
 amsmath,amssymb,
 aps,
]{revtex4-1}

\usepackage{multirow}
\usepackage{mathrsfs}
\usepackage{array,mathtools}
\usepackage{stmaryrd}
\usepackage{scrextend}
\usepackage[caption=false]{subfig}
\usepackage[normalem]{ulem}
\newcommand{\stkout}[1]{\ifmmode\text{\sout{\ensuremath{#1}}}\else\sout{#1}\fi}
\newcolumntype{C}{>{$}c<{$}}
\AtBeginDocument{
\heavyrulewidth=.08em
\lightrulewidth=.05em
\cmidrulewidth=.03em
\belowrulesep=.65ex
\belowbottomsep=0pt
\aboverulesep=.4ex
\abovetopsep=0pt
\cmidrulesep=\doublerulesep
\cmidrulekern=.5em
\defaultaddspace=.5em
}
\usepackage{booktabs}
\usepackage{graphicx}
\usepackage{dcolumn}
\usepackage{bm}
\usepackage{color}
\usepackage[dvipsnames]{xcolor}
\usepackage{notes2bib}

\begin{document}

\title{First-principles theory of infrared vibrational spectroscopy in metals and semimetals:\\ application to graphite}

\author{Luca Binci}
\altaffiliation{Current address: {\'E}cole Polytechnique F{\'e}d{\`e}rale de Lausanne (EPFL), Station 9, CH-1015 Lausanne, Switzerland.}
\affiliation{Dipartimento di Fisica, Universit{\`a} di Roma La Sapienza, Piazzale Aldo Moro 5, I-00185 Roma, Italy.}
\affiliation{Graphene Labs, Fondazione Istituto Italiano di Tecnologia, Via Morego, I-16163 Genova, Italy.}
\author{Paolo Barone}
\affiliation{%
SPIN-CNR, c/o Universit{\`a} G. D'Annunzio, I-66100 Chieti, Italy.}%
\author{Francesco Mauri}
\affiliation{Dipartimento di Fisica, Universit{\`a} di Roma La Sapienza, Piazzale Aldo Moro 5, I-00185 Roma, Italy.}
\affiliation{Graphene Labs, Fondazione Istituto Italiano di Tecnologia, Via Morego, I-16163 Genova, Italy.}

\date{\today}

\begin{abstract}
We develop an \emph{ab initio} method to simulate the infrared vibrational response of metallic systems in the framework of time-dependent density functional perturbation theory.
By introducing a generalized frequency-dependent Born effective charge tensor, we show that phonon peaks in the reflectivity of metals can be always described by a Fano function, whose shape is determined by the complex nature of the frequency-dependent effective charges and electronic dielectric tensor. The IR vibrational properties of graphite, chosen as a representative test case to benchmark our method, are found to be accurately reproduced. Our approach 
offers a first-principle scheme for the prediction and understanding of IR reflectance spectra of metals, that may represent one of the few available tools of investigation of these materials when subjected to extremely high-pressure conditions.
\end{abstract}
\maketitle
\section{introduction}
Infrared (IR) spectroscopy is a well-established technique for analyzing the vibrational properties of crystalline solids. In insulating or semiconducting materials, not displaying electronic intraband transitions in the IR region of the electromagnetic spectrum, phonon features arise at energies much smaller than the band gap, and hence they can be clearly identified. On the other hand, the IR response in metals is dominated by the Drude peak {-- the signature of the free electron response -- whose amplitude is proportional to the free carrier density $\rho$, and therefore to the square of the plasma frequency $\rho\propto\omega_\text{p}^2$}. The presence of a strong Drude peak generally precludes the detection of the vibrational features \cite{Bruesch1986phonons}. There are cases, however, in which this technique can also be useful in metallic materials, and even situations in which it is one of the only possible choices. {To better understand these cases, it is useful to recall the quantum-mechanical dependence of $\omega_\text{p}$ from the density of states at the Fermi level $D(E_{\text{F}})$ and the average electronic velocity at the Fermi surface $\bar{v}_\text{F}$: $\omega_\text{p}^2\propto\bar{v}^2_\text{F}\,D(E_{\text{F}})$. From this expression it is clear that, for example, systems with a pseudogap - like high-temperature superconducting cuprates \cite{Timusk1999} or transition metal dichalcogenides undergoing a charge-density wave transition \cite{Ruzicka2001,Borisenko2008,Umemoto2018} - displaying by definition a density of states which decreases significantly at the Fermi level, will have a small $\omega_\text{p}$; this condition narrows the Drude peak, and makes the vibrational features sharp enough to be investigated (see e.g. Fig. 1 of Ref. \cite{Baldassarre2008}, or Fig. 1 of of Ref. \cite{Moon2014})}. Another significant example is given by systems under extremely high pressure conditions, i.e. above hundreds of GPa, like H$_3$S \cite{Drozdov2015,Capitani2017}, LaH$_{10}$ \cite{Drozdov2019,Somayazulu2019} and the recently discovered semimetallic phase of hydrogen \cite{Eremets2019}. For these systems, the experiments in order to be performed require a setup composed of diamond anvil cells, and samples whose size is of the order of $\mu$m. For the determination of the crystal structure, the reduced sample size and the diamond-anvil environment -- together with the presence of a light element like hydrogen -- preclude the use of neutron scattering and makes very challenging an X-ray diffraction analysis \cite{Capitani2017,Ji2019}. Raman and IR spectroscopies represent alternative routes to this goal. However, {Raman spectroscopy of metals turns out to be difficult because of the smallness of the light penetration depth within the sample $\delta\propto(\omega\sigma_0)^{-1/2}$, for $\omega$ in the visible light and high dc conductivity $\sigma_0$. Therefore, owing to the lower frequencies employed,} IR reflectivity measurements represent one of the few possible and effective approach for investigating the crystal structure of metallic materials under extremely high-pressure conditions \cite{Capitani2017}.  
\section{theory}
\emph{Ab initio} calculations play a crucial role in the physical interpretation of experimental results. In this paper, by means of a time-dependent formulation of density functional perturbation theory (DFPT) \cite{Gross1985, Baroni2001}, we introduce a new method to simulate from first-principles the IR reflectivity absorption spectrum of metallic systems. 

The determination of the dielectric tensor $\bm{\epsilon}(\omega)$ gives a complete characterization of all the features appearing in IR spectra \cite{Born1988dynamical,Bistoni2019,Cappelluti2012}. {It can be decomposed as:}
\begin{equation}
\bm{\epsilon}(\omega) =\bm{\epsilon}^{\text{e}}(\omega)+4\pi\sum_s \bm{\chi}_s^{\text{I}}(\omega) \label{totepsilon}.
\end{equation}
where $\bm{\epsilon}^{\text{e}}(\omega)$ is the electronic dielectric tensor at fixed ions and $\bm{\chi}_s^{\text{I}}(\omega)$ represents the ionic contribution due to a phonon mode with index $s$
\begin{equation}
\bm{\chi}^{\text{I}}_s(\omega)=\frac{e^2}{\Omega}\,\frac{\textbf{d}_{s}(\omega_s)\,\textbf{d}_s(\omega_s)}{\omega^2_s-(\omega+i\gamma_s/2)^2}.
\label{ionicepsilon}
\end{equation}
Here, $\Omega$ is the unit cell volume, $\omega_s$ and $\gamma_s$ are the phonon frequency and inverse lifetime, respectively, and $\textbf{d}_s(\omega_s)=\sum_{\kappa}\bm{Z}_{\kappa}(\omega_s)\cdot\frac{\textbf{e}_{s,\kappa}}{\sqrt{m_\kappa}}~$ is the oscillator strength defined in terms of a frequency-dependent effective charge tensor $\bm{Z}(\omega_s)$, with $m_\kappa$ the mass of the $\kappa$-th atom and $\textbf{e}_{s,\kappa}$ the polarization vector of the $s$-th mode. {Eqs. ({\ref{totepsilon}}, \ref{ionicepsilon}) can be derived either by a phenomenologial approach \cite{Born1988dynamical}, or with more rigorous field-theoretic methods \cite{Bistoni2019,Cappelluti2012}}. Since phonon peaks are typically rather sharp, we can approximate $\bm{\epsilon}^\text{e}(\omega)$ by its value $\bm{\epsilon}^\text{e}_s=\bm{\epsilon}^\text{e}(\omega_s)$ at each given phonon mode $s$. 

Information from IR studies are generally extracted from transmission $T(\omega)$ and reflectivity $R(\omega)$ measurements. In the case of metals, $T(\omega)$ can be obtained only for very thin materials, whereas $R(\omega)$ can always be acquired. The standard formula of the reflectivity between the vacuum and the sample along one of the principal dieletric axes $\alpha$ is $R_{\alpha}(\omega)=\left|\frac{\sqrt{\epsilon_\alpha(\omega)}-1}{\sqrt{\epsilon_\alpha(\omega)}+1}\right|^2$ \cite{Bruesch1986phonons}, where $\epsilon_{\alpha}(\omega)$ is the diagonal element of the dielectric tensor. The general shape of the vibrational features in IR reflectivity spectra of metals can be deduced by a Taylor expansion of $R_{\alpha}(\omega)$ around a given $s$ phonon mode. In the limit $|\epsilon^{\text{e}}_{s,\alpha}|\gg|\chi_{s,\alpha}^{\text{I}}(\omega)|$ the reflectivity reads (see Appendix \ref{approxR}): 
\begin{equation}	
\begin{split}
R_{s,\alpha}(\omega)\sim R^{\text{e}}_{s,\alpha}\left[1+2\,\textsf{Re}\,\bigg(\frac{4\pi\chi^{\text{I}}_{s,\alpha}(\omega)}{\sqrt{\epsilon^{\text{e}}_{s,\alpha}}\,(\epsilon^{\text{e}}_{s,\alpha}-1)}\bigg)\right]\label{reflExp}.
\end{split}
\end{equation}
where $R^{\text{e}}_{s,\alpha}$ is the purely electronic reflectivity. The vibrational contribution (second term in the square parenthesis) can be recast as a Fano profile
\begin{equation}
\begin{split}
\textsf{Re}\left(\frac{4\pi\chi^{\text{I}}_{s,\alpha}(\omega)}{\sqrt{\epsilon^{\text{e}}_{s,\alpha}}\,(\epsilon^{\text{e}}_{s,\alpha}-1)}\right)=W_{s,\alpha}\,\frac{q_{s,\alpha}^2-1+2q_{s,\alpha}\,\xi_s(\omega)}{(1+q_{s,\alpha}^2)\big(1+\xi^2_s(\omega)\big)}.
\end{split}\label{Fano}
\end{equation}
Here we defined the following quantities:
\begin{gather}
W_{s,\alpha}=\frac{\big|D_{s,\alpha}\big|^2}{\gamma_s\omega_s};\qquad q_{s,\alpha}=-\frac{\textsf{Re}\,D_{s,\alpha}}{\textsf{Im}\,D_{s,\alpha}}\label{asymmetry}\\
(D_{s,\alpha})^2=i\,\frac{4\pi e^2}{\Omega}\frac{(d_{s,\alpha})^2}{\sqrt{\epsilon^{\text{e}}_{s,\alpha}}\,(\epsilon^{\text{e}}_{s,\alpha}-1)}\label{oscStrenght}
\end{gather}	 
{whereas} $\xi_s(\omega)=(\omega^2-\omega_s^2)/\gamma_s\omega$, which, close to a phonon peak, can be approximated to $\xi_s(\omega)\sim 2(\omega-\omega_s)/\gamma_s$. Eq.(\ref{Fano}) is a Fano function in the variable $\xi_s(\omega)$ \citep{Fano1961} which is completely determined by five parameters, namely $R^{\text{e}}_s$, $\omega_s$, $\gamma_s$, $W_{s,\alpha}$, $q_{s,\alpha}$. This implies that, in the case of metals, it cannot be used to obtain all the six parameters characterizing $\bm{\epsilon}(\omega)$, i.e. $(\textsf{Re} \ \bm{Z}^{\text{e}}_\kappa,\textsf{Im} \ \bm{Z}^{\text{e}}_\kappa)$, $(\textsf{Re} \ \bm{\epsilon}_s^{\text{e}},\textsf{Im} \ \bm{\epsilon}_s^{\text{e}})$, $\gamma_s$, $\omega_s$ \cite{Nemanich1977, Capitani2017}. Instead, the proper way to obtain both the real and the imaginary parts of $\bm{Z}^{\text{e}}_\kappa$ is to fit the vibrational contribution of the real part of the optical conductivity; {this can be done once the smooth electronic part has been previously subtracted}, as it was done e.g. in Ref. \cite{Kuzmenko2009}. {The reflectivity expansion in Eqs. (\ref{reflExp}, \ref{Fano}) holds for every material, both metallic and insulating. Indeed, for $\mathsf{Im}\,\bm{Z}^{\mathrm{e}}_\kappa\rightarrow \mathbf{0}$ ($q_{s,\alpha}\rightarrow-\infty$) one recovers the standard insulating limit of a Lorentzian shape. For metals instead, as static polarisation is not a well-defined quantity, the effective charge can only be defined in the dynamical ($\omega$-dependent) version. We show that its imaginary part is the responsible of the Fano shape of phonon peaks in reflectivity spectra.}

In DFPT, both the $\bm{\epsilon}^{\text{e}}$ and $\bm{Z}_\kappa^{\text{e}}$ are defined as derivatives of the electronic polarization, the former with respect to the electric field $\textbf{E}$, the latter with respect to the ionic displacements $\textbf{u}_\kappa$ \cite{Gonze1997}. The effective charge tensor can be decomposed in two contributions $\bm{{Z}}_\kappa=\textbf{1}Z^{\text{I}}_\kappa+\bm{Z}^{\text{e}}_\kappa$,  the first (constant) term $Z^{\text{I}}_\kappa$ is the (pseudo)charge of the nuclei, while the second electronic contribution $\bm{Z}^{\text{e}}_\kappa$ is due to the interaction between the electrons and the lattice; in the following we will focus on this last one. 
Within time-dependent DFPT \cite{Gross1985, Baroni2001} and adopting the variational approach proposed in \cite{Calandra2010}, the effective charge tensor can be expressed as:
\begin{equation}
\begin{split}
&e\bm{Z}_\kappa^{\text{e}}\big[n^\textbf{E},n^{\textbf{u}_\kappa}\big](\omega_s)= -\frac{2}{N_\textbf{k}}\sum_{\textbf{k},nm}\frac{f_{\textbf{k},n}-f_{\textbf{k},m}}{ {(E_{\textbf{k},n}-E_{\textbf{k},m})^2-z_s^2}} \\
&\times  \Big\langle u_{\textbf{k},m}\Big|\Big( i e \hbar\textbf{v}_\textbf{k} +(E_{\textbf{k},n}-E_{\textbf{k},m}) {V}^{\textbf{E}}_{\text{Hxc}}\Big) \Big|u_{\textbf{k},n}\Big\rangle\\
&\times\Big\langle u_{\textbf{k},n}\Big|\Big({V}_{\text{I}}^{\textbf{u}_\kappa}+{V}_{\text{Hxc}}^{\textbf{u}_\kappa} \Big)\Big|u_{\textbf{k},m}\Big\rangle\\
&+\int \text{d}^3r\,\text{d}^3r'\,n^{\textbf{E}}(\textbf{r},\omega_s)\,K_{\text{Hxc}}(\textbf{r},\textbf{r}')\,n^{\textbf{u}_\kappa}(\textbf{r}',\omega_s)\label{effCharge},
\end{split}
\end{equation}
while the electronic dielectric tensor $\bm{\epsilon}^{\text{e}}(\omega_s)$ reads:
\begin{gather}
\begin{split}
&\bm{\epsilon}^{\text{e}}\big[n^\textbf{E}\big](\omega_s)=\textbf{1}+4\pi\\
&\times\frac{2}{N_\textbf{k}\Omega}\sum_{\textbf{k},nm}\frac{1}{ {(E_{\textbf{k},n}-E_{\textbf{k},m})^2-z_s^2}}\frac{f_{\textbf{k},n}-f_{\textbf{k},m}}{E_{\textbf{k},n}-E_{\textbf{k},m}}\\
&\times\Big\langle u_{\textbf{k},m}\Big|\Big( i e \hbar\textbf{v}_\textbf{k} +(E_{\textbf{k},n}-E_{\textbf{k},m}) {V}^{\textbf{E}}_{\text{Hxc}}\Big)\Big|u_{\textbf{k},n} \Big\rangle\\
&\times \Big\langle u_{\textbf{k},n}\Big|\Big( i e \hbar\textbf{v}_\textbf{k} +
(E_{\textbf{k},m}-E_{\textbf{k},n}) {V}^{\textbf{E}}_{\text{Hxc}}\Big) \Big|u_{\textbf{k},m} \Big\rangle\\
&+\frac{4\pi}{\Omega}\int\text{d}^3r\,\text{d}^3r'\,n^{\textbf{E}}(\textbf{r},\omega_s)\,K_{\text{Hxc}}(\textbf{r},\textbf{r}')\,n^{\textbf{E}}(\textbf{r}',\omega_s).\label{eleSusc}
\end{split}
\end{gather}
In these expressions, $N_\textbf{k}$ is the number of points in the \textbf{k}-grid, $\textbf{v}_\textbf{k}=\frac{1}{\hbar}\frac{\partial H^0_\textbf{k}}{\partial \textbf{k}}  $, where $H^0_\textbf{k}=e^{-i\textbf{k}\cdot\textbf{r}}H^0 e^{i\textbf{k}\cdot\textbf{r}}$ and $H^0$ is the unperturbed Kohn-Sham (KS) Hamiltonian; $E_{\textbf{k},n}$ is the unperturbed KS eigenvalue and $u_{\textbf{k},n}$ is the periodic part of the {corresponding} KS eigenstate in the Bloch form \cite{Kohn1965}. We denote with $f_{\textbf{k},n}$ the smearing function, while $z_s=\hbar\omega_s+i\eta_s$, where $\eta_s$ is a positive small real number with the dimension of an energy. The frequency-dependent charge density induced by $\bm{\xi}=\textbf{E},\textbf{u}_\kappa$, denoted as $n^{\bm{\xi}}=\frac{\partial n}{\partial\bm{\xi}}$, gives rise to a Hartree and exchange-correlation (Hxc) potential:
\begin{equation}
{V}_{\text{Hxc}}^{\bm{\xi}}\big[n^{\bm{\xi}}\big]({\textbf{r}},\omega_s)=\int \text{d}^3r\,K_{\text{Hxc}}(\textbf{r},\textbf{r}')\,n^{\bm{\xi}}(\textbf{r}',\omega_s),
\end{equation}
where $K_{\text{Hxc}}(\textbf{r},\textbf{r}')=\frac{\delta^2 E_{\text{Hxc}}[n]}{\delta n(\textbf{r})\delta n(\textbf{r}')}$ is the Hxc kernel. Finally, the first-order perturbative expressions of the induced charge density with respect to the electric field and ionic displacements are:
\begin{equation}
\begin{split}
&n^{\textbf{E}}(\omega_s)= \frac{2}{N_\textbf{k}}\sum_{\textbf{k},nm}\frac{f_{\textbf{k},n}-f_{\textbf{k},m}}{ {(E_{\textbf{k},n}-E_{\textbf{k},m})^2-z_s^2}}\,u^\ast_{\textbf{k},n}\,u_{\textbf{k},m}\\
&\times \Big\langle u_{\textbf{k},m}\Big|\Big( i e \hbar\textbf{v}_\textbf{k} +(E_{\textbf{k},n}-E_{\textbf{k},m}) {V}^{\textbf{E}}_{\text{Hxc}} \Big)\Big|u_{\textbf{k},n} \Big\rangle\label{n1E},
\end{split}
\end{equation}
and
\begin{equation}
\begin{split}
&\!n^{\textbf{u}_\kappa}(\omega_s)=\frac{2}{N_\textbf{k}}\sum_{\textbf{k},nm}\frac{f_{\textbf{k},n}-f_{\textbf{k},m}}{ {(E_{\textbf{k},n}-E_{\textbf{k},m})^2-z_s^2}}\,u^\ast_{\textbf{k},n}\,u_{\textbf{k},m}\\
&\times(E_{\textbf{k},n}-E_{\textbf{k},m})\,\Big\langle u_{\textbf{k},m}\Big|\Big({V}_{\text{I}}^{\textbf{u}_\kappa}+{V}_{\text{Hxc}}^{\textbf{u}_\kappa}\Big)\Big|u_{\textbf{k},n}\Big\rangle.\label{n1u}
\end{split}
\end{equation}
To highlight certain analytical properties, we wrote the previous expressions in a slightly different way with respect to their standard form \cite{Calandra2010}. It is clear that only $\lim_{z\rightarrow0}n^{\textbf{u}_\kappa}$ is in general well-defined: in the limit of infinite $\textbf{k}$ points, all the other quantities have  an integrand which can become arbitrarily large when $z\rightarrow0$ if the Fermi level falls between the intersection of two bands $(\tilde{m},\tilde{n})$ at a point $\textbf{k}^\ast$, where $\lim_{\textbf{k}\rightarrow\textbf{k}^\ast} E_{\textbf{k},\tilde{n}}-E_{\textbf{k},\tilde{m}} \rightarrow0$. Note that -- contrary to the induced charge densities Eqs. (\ref{n1E}, \ref{n1u}) and the effective charge tensor Eq.(\ref{effCharge}) -- only $\bm{\epsilon}^{\text{e}}$ contains a contribution from intraband terms, which is always divergent when $z\rightarrow0$ and gives rise to the Drude peak. It is therefore necessary to mantain the frequency dependence of all relevant quantities -- namely, $n^{\bm{\xi}}$, $\bm{Z}^{\text{e}}_\kappa$ and $\bm{\epsilon}^{\text{e}}$ -- in order to have a stable and robust implementation of Eqs. (\ref{effCharge}, \ref{eleSusc}, \ref{n1E}) in a first-principles code.

Although feasible in principle, a full implementation would require a substantial rewriting of the linear response code. Therefore, for computational semplicity, we exploit the variational property with respect to the first-order charge density of Eqs. (\ref{effCharge}, \ref{eleSusc}) (as discussed in Ref. \cite{Calandra2010}), neglecting in all the self-consistent loops the imaginary part of the first order charge density:
\begin{gather}
\bm{Z}^{\text{e}}_\kappa(\omega_s)\approx\bm{Z}^{\text{e}}_\kappa\big[\textsf{Re}\,n^\textbf{E},\textsf{Re}\,n^{\textbf{u}_\kappa}\big](\omega_s)\label{ZApprox}\\
\bm{\epsilon}^{\text{e}}(\omega_s)\approx\bm{\epsilon}^{\text{e}}\big[\textsf{Re}\,n^\textbf{E}\big](\omega_s).\label{ChiApprox}
\end{gather}
Regarding the practical implementation in the code, we used a dynamical extension of the linear response formalism described in Ref. \cite{Baroni2001}, which is equivalent to the one of Ref. \cite{Calandra2010}, employing a frequency-dependent Sternheimer equation with a similar scheme of Ref. {\cite{Giustino2010}} (see Appendix \ref{implem} for more details). 
\section{applications}
\subsection{Graphite}
We benchmark our approach by evaluating the reflectivity spectra of bulk graphite and analyzing the IR peaks $E_{1u}$ ($\omega_{E_{1u}}=$ 1587 cm$^{-1}$) and $A_{2u}$ ($\omega_{A_{2u}}=$ 868 cm$^{-1}$), which have been thoroughly investigated by IR spectroscopy measurements \cite{Manzardo2012,Nemanich1977,Underhill1979,Leung1980,Philipp1977,Kuzmenko2008,Papoular2014,Venghaus1977,Venghaus1975,Draine2016}. Lattice symmetry forces tensorial quantities to be diagonal, and the in-plane elements to be equal. In the following, we will denote as $T_\parallel$ ($T_\perp$) the components of a given tensor $\bm T$ parallel (perpendicular) to the graphene sheets. \emph{Ab initio} calculations were performed using the PW and PHonon packages of Quantum ESPRESSO \cite{Giannozzi2009,Giannozzi2017}, within which we implemented the theory described above. We use local-density approximation \cite{Ceperley1980}, norm-conserving pseudopotentials \cite{Troullier1993}, Fermi-Dirac smearing and a plane-wave expansion up to 55 Ry cutoff. We choose the value of $\eta_s$ to be one order of magnitude smaller than the corresponding $\omega_s$ values: $\eta_{E_{1u}}=110$ cm$^{-1}$ and $\eta_{A_{2u}}=55$ cm$^{-1}$. We account for the termal expansion using as lattice constant $c=6.68\;\text{\AA}$ for $T=150$ K and $c=6.70\;\text{\AA}$ for $T=300$ K, while keeping the in-plane lattice constant fixed at $a=2.46\;\text{\AA}$ \cite{Boettger1997}. We use as $\omega_s$ the values taken from Ref. \cite{Nemanich1977}. Regarding the \textbf{k}-point sampling, as graphene, bulk graphite is a semimetal with valence and conduction bands touching and crossing at the high-symmetry \textbf{K} and \textbf{H} points in the Brillouin zone, so interband electronic transitions will contribute to the optical response for any value of $\hbar\omega$. At IR frequencies, the most significant contributions to the $E_{1u}$ absorption come from a small cylinder along the \textbf{K}-\textbf{H} line, in which the denominators in Eqs. (\ref{effCharge}, \ref{eleSusc}) reach their minimum value. Thus, a very fine \textbf{k}-points grid around this region is needed for an accurate evaluation of the sum over $\textbf{k}$ appearing in Eqs. (\ref{effCharge}, \ref{eleSusc}). To this end, we employed a mesh with a uniform sampling along the $k_z$ direction, and a non-uniform grid within the $(k_x,k_y)$ plane, where the density of \textbf{k}-points increases exponentially around the \textbf{K}-\textbf{H} line obeying a $C_3$ symmetry (details can be found in Appendix \ref{kpoints}). 

\begin{figure}[]
\centering
\includegraphics[width=8cm]{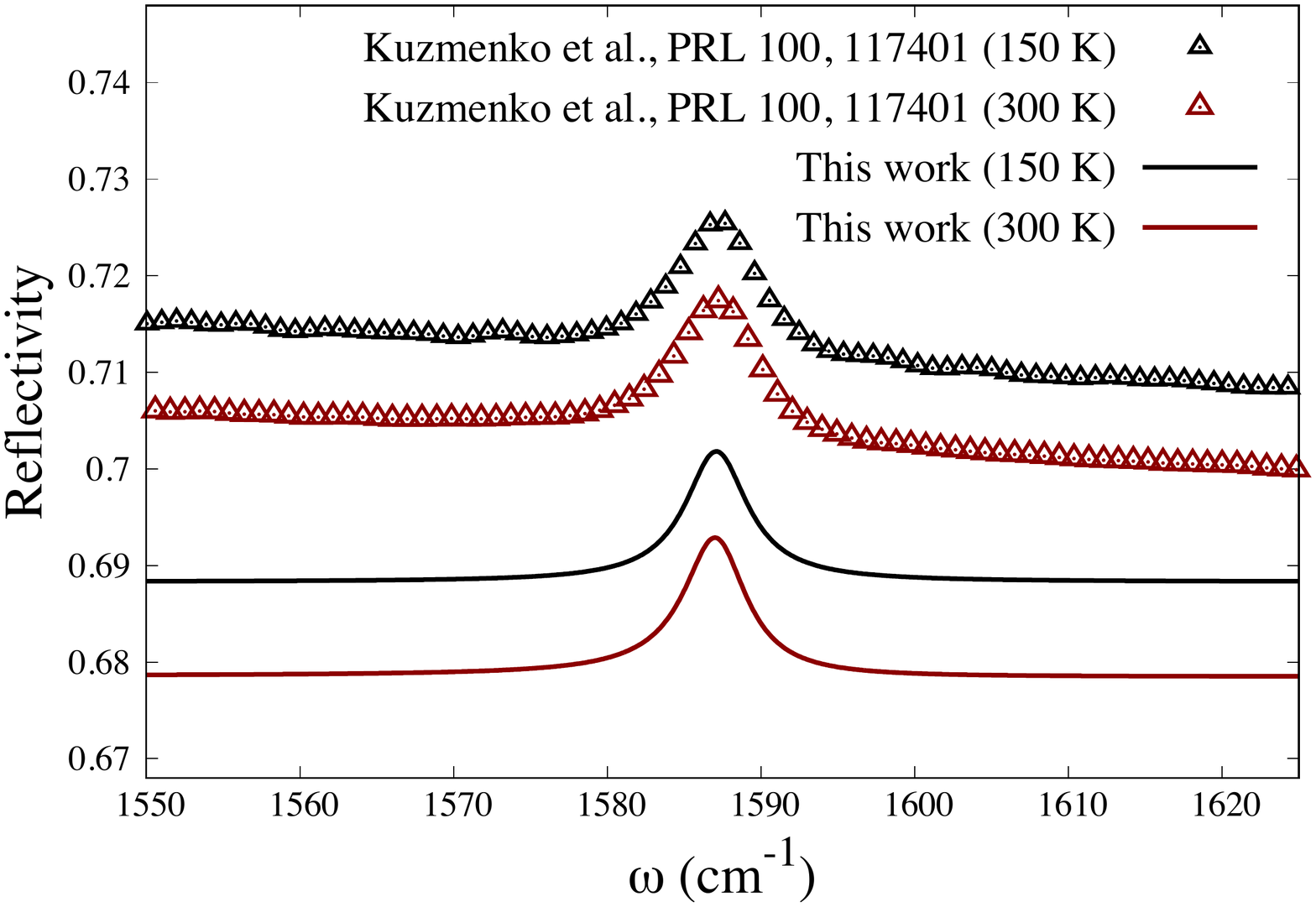}
\caption{Simulated (lines) and experimental (dots) phonon peak associated to the $E_{1u}$ mode. We shifted the experimental peaks so that their tips are at 1587 cm$^{-1}$.}\label{2Fig}
\end{figure}

\begin{figure}[h]
\centering
\includegraphics[width=8cm]{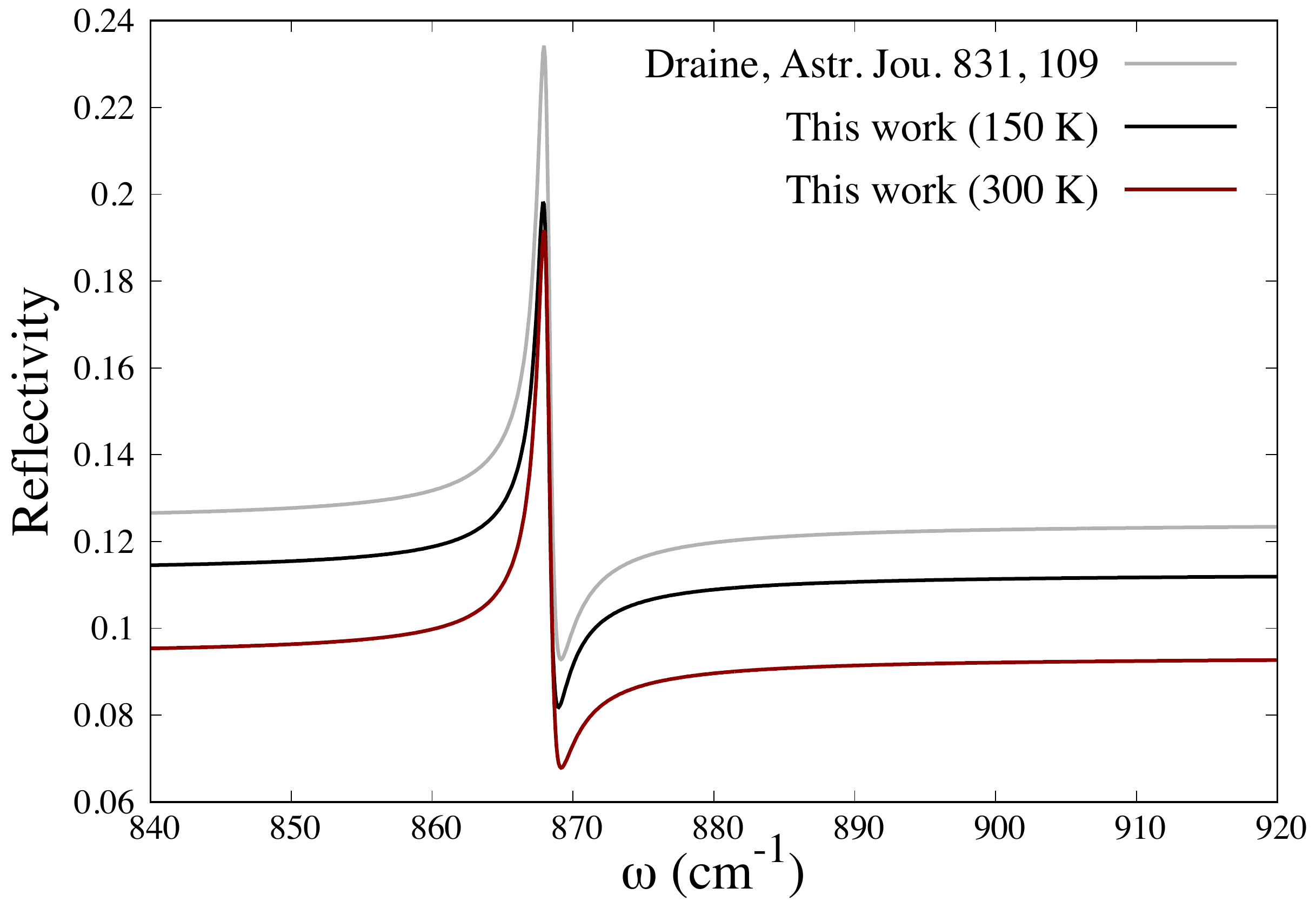}
\caption{Comparison between our \emph{ab initio} results (black and red lines) and a fit (whose paramenters are not related to a particular $T$) taken from Ref. \cite{Draine2016} (grey line) of the reflectivity around the resonance of the $A_{2u}$ mode.}\label{3Fig}
\end{figure}

In Fig. \ref{2Fig} and Fig. \ref{3Fig} we compare our simulated reflectivity with the experimental data. Importantly, we notice that the symmetry of the peaks depends on the phases of both $\bm{Z}_\kappa(\omega_s)$ and $\bm{\epsilon}^{\text{e}}(\omega_s)$. In fact, from Eq. (\ref{oscStrenght}), $D_{s,\alpha}=|D_{s,\alpha}|e^{i\arg D_{s,\alpha}}$ where {for $|\epsilon_{s,\alpha}^{\text{e}}|\gg1$} $\arg D_{s,\alpha}\approx \frac{\pi}{4} + \arg d_{s,\alpha} -\frac{3}{4}\arg\epsilon^{\text{e}}_{s,\alpha}$ and, from Eq. (\ref{asymmetry}), $\tan\,(\arg D_{s,\alpha})=-\frac{1}{q_{s,\alpha}}$. Using {our computed values in Tables \ref{tab1} and \ref{tab2}, for the $E_{1u}$ peak (Fig. \ref{2Fig}) we find $q_{E_{1u},\parallel}\approx -52\,(-19)$ at $T=150\,(300)$ K ($q\ll-1$ is the Lorentzian limit)}, which explains the symmetric shape of the resonance. Remarkably, the temperature dependence obtained from our calculations -- mainly due to the temperature dependence of $\bm{\epsilon}^{\text{e}}(\omega_s)$ -- well reproduces the one reported in Ref. \cite{Kuzmenko2008}. As for the $A_{2u}$ peak, we compare the calculated $R_\perp(\omega)$ with a fit proposed in Ref. \cite{Draine2016} and realized taking into account several experiments. As shown in Fig. \ref{3Fig}, also in this case our approach successfully reproduces the expected Fano asymmetric shape of the phonon peak, for which {$q_{A_{2u},\perp}\approx -1.3\,(-1.4)$ at $T=150 \,(300)$ K ($q=-1$ is the complete asymmetric case)}.

{We report in Table \ref{tab1} the oscillator strenghts $d_{\perp}$ $(d_{\parallel})$ of Eq. (\ref{ionicepsilon}) evaluated at $\omega_{E_{1u}}$ ($\omega_{A_{2u}}$) and properly rescaled by the square root of the carbon mass $m_\text{C}$ (and a factor 2 stemming from the scalar product with the polarization vectors $\textbf{e}_{s,\kappa}$), and we compare them with available experimental estimates. In the case of graphite, they are equal to the average of the absolute value of $Z_\parallel$ ($Z_\perp$) of the four C atoms of the unit cell}, that, because of symmetry, are exactly equal in pairs, with opposite sign. In addition, it is well-known that the components of the effective charge tensor obey the acoustic sum rule (ASR) $\sum_\kappa \bm{Z}_\kappa=\textbf{0}$ \cite{Born1988dynamical}; noteworthy, we found that this rule in general is not respected in the dynamical case (see Table \ref{tab1appendix} in Appendix \ref{depEta}). 

\begin{table}[]
\centering
\caption{Comparison between \emph{ab initio} and experimental{/theoretical} oscillator strenght $\tilde{\textbf{d}}_s=\sqrt{m_{\text{C}}/4}\,\textbf{d}_s$. The parallel component refers to the $E_{1u}$ mode, the perpendicular one to the $A_{2u}$ mode.}\label{tab1}
\begin{tabular}{*{6}{c}}
\midrule
\midrule
Reference& $T$ (K) &$\textsf{Re}\,\tilde{d}_\parallel$& $\textsf{Im}\,\tilde{d}_\parallel$ &$\textsf{Re}\,\tilde{d}_\perp$  &$\textsf{Im}\,\tilde{d}_\perp$  \\
\midrule
 Ref. \cite{Nemanich1977}\footref{note1}   & &   0.41 &  & 0.08 &   \\
 Ref. \cite{Underhill1979}\footref{note1} & 300 &0.18  &  &  &  \\
 Ref. \cite{Leung1980}\footref{note1} & 300& 0.21 &&&\\
  {Ref. \cite{Jeon2005}\footref{note2}}&&{0.014}&{0.015}&&\\
 \multirow{ 2}{*}{{Ref. \cite{Manzardo2012}\footnote{\label{note2} {theoretical}}}}&{150}&{0.17}&{0.15}&&\\
   &  {300} &{0.18}&{0.15}  &  &\\
   \multirow{ 2}{*}{Ref. \cite{Manzardo2012}\footnote{\label{note1} experimental}} &  150 &0.29&0.14  &  &\\
   &  300 &0.31&0.13  &  &\\
 \midrule
Present work & 150 &  0.27 & 0.09 & 0.07 &  {0.0001}\\
Present work & 300 &  0.27 & 0.10 & 0.07 &  {0.0001}\\
\midrule
\midrule
\end{tabular}
\end{table}

We stress the fact that in older experiments the effective charge was supposed to be a real quantity and the imaginary part was completely neglected \cite{Nemanich1977,Underhill1979, Leung1980}, an assumption that has been relaxed only recently \cite{Manzardo2012}. Such neglect of the complex nature of the effective charges may have {led} to inaccurate results because of a wrong fitting procedure of the experimental data. Within our approach, we find that indeed the imaginary part of $d_\parallel$ is approximatively 1/3 of the real part, in good agreement with the experimental results reported in Ref. \cite{Manzardo2012}. The out-of-plane component $d_\perp$ has a negligible imaginary part, whereas the real one is found to be one order of magnitude smaller than ${\textsf{Re}\,}d_\parallel$ (reflecting the dielectric-like properties of graphite in the transverse direction), with our calculations yielding a value in excellent agreement with the experimental estimates \cite{Nemanich1977}. We find that the combined effect of the thermal lattice expansion and the increase of electronic temperature has no effect on $\textsf{Re}\,d_\parallel$ and $d_\perp$. Instead, $\textsf{Im}\,d_\parallel$ does not depend appreciably on thermal expansion, but it increases with $T$, contrary to Ref. \cite{Manzardo2012}. However, it is worth to mention that the parameter $\eta_s=\textsf{Im}\,z_s$, accounting for the damping of the electronic states, is also $T$-dependent. We have not studied such dependence; nevertheless in Appendix \ref{depEta} we show that the variation of $\eta_s$ can markedly affects $\textsf{Im}\,{d}_\parallel$.

\begin{table}[]
\centering
\caption{Comparison between \emph{ab initio} and experimental dielectric tensor. For comparison, we also report the value of $\textsf{Re}\,\epsilon_\perp$ at $\omega=9679$ cm$^{-1}$ (taken from Ref. \cite{Venghaus1975}) because, although it is not the characteristic frequency of the $A_{2u}$ mode, the dielectric tensor is not expected to vary appreciably from 686 to 9679 cm$^{-1}$ since there are no interband electronic transitions in such frequency range.}\label{tab2}
\begin{tabular}{*{6}{c}}
\midrule
\midrule
Reference & $T$ (K) &$\textsf{Re}\,\epsilon_\parallel$ &$\textsf{Im}\,\epsilon_\parallel$&$\textsf{Re}\,\epsilon_\perp$&$\textsf{Im}\,\epsilon_\perp$  \\
\midrule
 Ref. \cite{Philipp1977}\footref{note3} &   &8.8&50&&\\
 Ref. \cite{Papoular2014}\footnote{\label{note3} as reported by Ref. \cite{Draine2016}} & &0.73&73&&\\
 Ref. \cite{Venghaus1977}\footref{note3} & &&&5.3&0.68\\
 Ref. \cite{Venghaus1975}\footref{note3}& &&&3.3&\\
 Ref. \cite{Draine2016}\footnote{values corresponding to a fit realized taking into account many experimental data.}& &&&4.2&0.89\\
 \midrule
Present work&150 &6.1&62&3.9&0.79\\
Present work&300 &7.9&59&3.4&0.71\\
\midrule
\midrule
\end{tabular}
\end{table}

As for the electronic dielectric tensor $\bm{\epsilon}^{\text{e}}(\omega_s)$, in Table \ref{tab2} we report $\epsilon_\parallel$ ($\epsilon_\perp$) evaluated at $\omega_{E_{1u}}$ ($\omega_{A_{2u}}$). We found that the components of the dielectric tensor depend only on the electronic temperature and not on the lattice thermal expansion. Also for $\bm{\epsilon}^{\text{e}}(\omega_s)$, as well as for $\bm{Z}_\kappa(\omega_s)$, the in-plane components are always larger than the out-of-plane ones, both for the real and the imaginary parts. This is a consequence of the mirror reflection symmetry with respect to the carbon planes, that is exact in monolayers like graphene, and only approximate in graphite, which forbides low-energy electronic excitations for perturbations perpendicular to the layered structure in the linear response regime. 

We finally mention that we also computed the nonadiabatic/adiabatic phonon frequencies for graphite (see next section), finding $\omega_{E_{2g}}^\mathrm{NA(A)}=1560.8\,(1560.5)$ cm$^{-1}$ and $\omega_{E_{1u}}^\mathrm{NA(A)}=1567.8\,(1569.4)$ cm$^{-1}$ at $T=300$ K; this gives a nonadiabatic(adiabatic) splitting of 7(9) cm$^{-1}$, which well compares with the experimental splitting of $7$ cm$^{-1}$ at the same temperature \cite{Giura2012}.

\subsection{Graphene bilayer and trilayers}
As additional benchmarks, in this part we discuss the application of our theory to two related systems: graphene bilayer and the two types of graphene trilayer, with Bernal (ABA) and rhombohedral (ABC) stacking. Since, to the best of our knowledge, direct reflectivity experimental data are currently not available for  these materials, we focus here on the effective charge tensor. Moreover, together with the oscillator strenghts $\mathbf{d}_s$, we have also calculated both the adiabatic ($\omega_\mathrm{A}$) and nonadiabatic ($\omega_\mathrm{NA}$) optical phonon frequencies at $\mathbf{q}=\bm{\Gamma}$. To perform these last calculations, we modified the \texttt{ph.x} code of Quantum ESPRESSO, in order to compute $\omega_{\mathrm{NA}}$ at the $\bm{\Gamma}$ point of the BZ. The specific form of the finite-frequency generalized dynamical matrix at $\mathbf{q}=\bm{\Gamma}$ is analogous to Eq.(\ref{effCharge}), with the electric-field matrix elements replaced by the electron-phonon ones, and is obtained using the same variational formulation and the same approximation of the charge density (i.e. the neglect of the imaginary part of the induced charge density) as for the effective charge and the dielectric tensors. This represents the same formulation as in Ref. \cite{Calandra2010}, but at variance with that, we retained the frequency dependence in the first-order charge density, and we calculate explicitly the double counting term (see the second term of Eq.(\ref{effCharge2}) in the appendix).    

At variance with graphite, a splitting of in-plane and out-of-plane optical modes occurs in trilayer graphene. For the ABC stacking, such splitting involves only the Raman-active modes $E_{2g}$ and $B_{1g}$: $E_{2g}\rightarrow E^{(1)}_g,E^{(2)}_g$ for the in-plane mode, and $B_{1g}\rightarrow A_{1g}^{(1)}, A_{1g}^{(2)}$ for the out-of-plane mode. For the ABA stacking instead, the splitting affects the IR-active modes $E_{2u}$ and $A_{2u}$: $E_{2u}\rightarrow E'_{(1)}, E'_{(2)}$, and $A_{2u}\rightarrow A''_{(1)},A''_{(2)}$ for the out-of-plane mode. The $E'_{(1)}, E'_{(2)}$ modes of ABA trilayer graphene are also Raman-active. For the bilayer, the splittings are analogous to the ones of graphite. 

In Tables \ref{bilayer} and \ref{trilayer} we show the results of our calculations. The temperature was set to 300 K for all the simulations, and we used the same lattice parameter and interlayer spacings of graphite at the same temperature. The adaptive \textbf{k}-point grid with parameters $(L,l,\mathcal{N})=(10,4,25)$ and 1 $N_{k_z}$ was used (see Appendix \ref{kpoints}), and all other computational details (pseudopotential, cutoff, etc.) are the same as for graphite.

\begin{table}[h]
\centering
\caption{Bilayer\label{bilayer}. The oscillator strenghts are $\tilde{\mathbf{d}}_s=\sqrt{m_{\mathrm{C}}/4}\,\mathbf{d}_s$}
\begin{tabular}{*{7}{c}}
\midrule
\midrule
&$E_g$\footnote{\label{ra}Raman-active}&$E_u$\footnote{\label{ir}IR-active}&$A_g$\footref{ra}&$A_u$\footref{ir}&exp \cite{Kuzmenko2009}&Theor \cite{Kuzmenko2009}\\
\midrule
$\textsf{Re}\,\tilde{d}_\parallel$&&0.211&&&$0.337\pm0.113$&0\\
$\textsf{Im}\,\tilde{d}_\parallel$&&0.104&&&&\\
$q_\parallel$&&-2.029&&&$-0.788\pm0.298$&--\\
\midrule
$\textsf{Re}\,\tilde{d}_\perp$&&&&0.013&&\\
$\textsf{Im}\,\tilde{d}_\perp$&&&&-0.001&&\\
\midrule
$\omega_{\text{A}}$ (cm$^{-1}$)&1561&1567&892&894&&\\
$\omega_{\text{NA}}$ (cm$^{-1}$)&1560&1565&892&894&&\\
\midrule
\midrule
\end{tabular}
\end{table}

For the bilayer graphene, we compare our results with the experimental and theoretical data of Ref. \cite{Kuzmenko2008}. Our results, although not precise as for graphite, yet represent a considerable improvement compared to the theoretical model employed in the same work. We point out that the quantitative discrepancy with the experimental results may also arise from the technical difficulties inherent to reflectivity measurements in bottom-gated bilayer graphene. Moreover, with our method, we are also able to calculate the out-of-plane oscillator strength, a quantity that to our knowledge has never been computed for multilayered graphene. We have also evaluated the static limit of the effective charge tensor for the bilayer, yielding a value $\lim_{\omega\rightarrow 0}\tilde{d}_{\parallel}(\omega)=0.349$. This is in good agreement with a theoretical value of 0.394 \cite{Bistoni2019}, obtained as averaged limit of zero electric field perpendicular to the graphene planes. Our calculated in-plane nonadiabatic phonon frequencies are very close to the adiabatic ones, in close analogy with the negligible nonadiabatic effect found in pristine graphite \cite{Saitta2008}. For the out-of-plane phonon frequencies, the inclusion of the $\omega$-dependence induces a variation smaller than cm$^{-1}$, that is in fact pratically negligible. 

\begin{table}[h]
\centering
\caption{Trilayer Bernal ABA (top) and rhombohedral ABC\label{trilayer} (bottom). The oscillator strenghts are $\tilde{\mathbf{d}}_s=\sqrt{m_{\mathrm{C}}/6}\,\mathbf{d}_s$}
\begin{tabular}{*{7}{c}}
\midrule
\midrule
\multicolumn{7}{c}{Bernal (ABA)}\\
&$E'_{(1)}$\footref{ra}\footref{ir}&$E'_{(2)}$\footref{ra}\footref{ir}&$E''$\footref{ra}&$A''_{(1)}$\footref{ir}&$A'$\footref{ra}&$A''_{(2)}$\footref{ir}\\
\midrule
$\textsf{Re}\,\tilde{d}_\parallel$&0.023&0.242&&&&\\
$\textsf{Im}\,\tilde{d}_\parallel$&0.162&0.118&&&&\\
\midrule
$\textsf{Re}\,\tilde{d}_\perp$&&&&0.013&&0.015\\
$\textsf{Im}\,\tilde{d}_\perp$&&&&-0.002&&-0.003\\
\midrule
$\omega_{\text{A}}$ (cm$^{-1}$)&1563&1564&1570&888&893&893\\
$\omega_{\text{NA}}$ (cm$^{-1}$)&1563&1566&1569&888&893&893\\
\midrule
\midrule\\
\midrule
\midrule
\multicolumn{7}{c}{Rhombohedral (ABC)}\\
&$E^{(1)}_g$\footnote{\label{ra}Raman-active}&$E_u$\footnote{\label{ir}IR-active}&$E^{(2)}_g$\footref{ra}&$A_g^{(1)}$\footref{ra}&$A_u$\footref{ir}&$A_g^{(2)}$\footref{ra}\\
\midrule
$\textsf{Re}\,\tilde{d}_\parallel$&&0.207&&&&\\
$\textsf{Im}\,\tilde{d}_\parallel$&&0.055&&&&\\
\midrule
$\textsf{Re}\,\tilde{d}_\perp$&&&&&0.011&\\
$\textsf{Im}\,\tilde{d}_\perp$&&&&&-0.002&\\
\midrule
$\omega_{\text{A}}$ (cm$^{-1}$)&1563&1567&1571&888&893&893\\
$\omega_{\text{NA}}$ (cm$^{-1}$)&1561&1563&1570&888&893&893\\
\midrule
\midrule
\end{tabular}
\end{table}

\begin{figure}[h]
\centering
\subfloat[][Bernal (ABA) stacking]{\includegraphics[width=8cm]{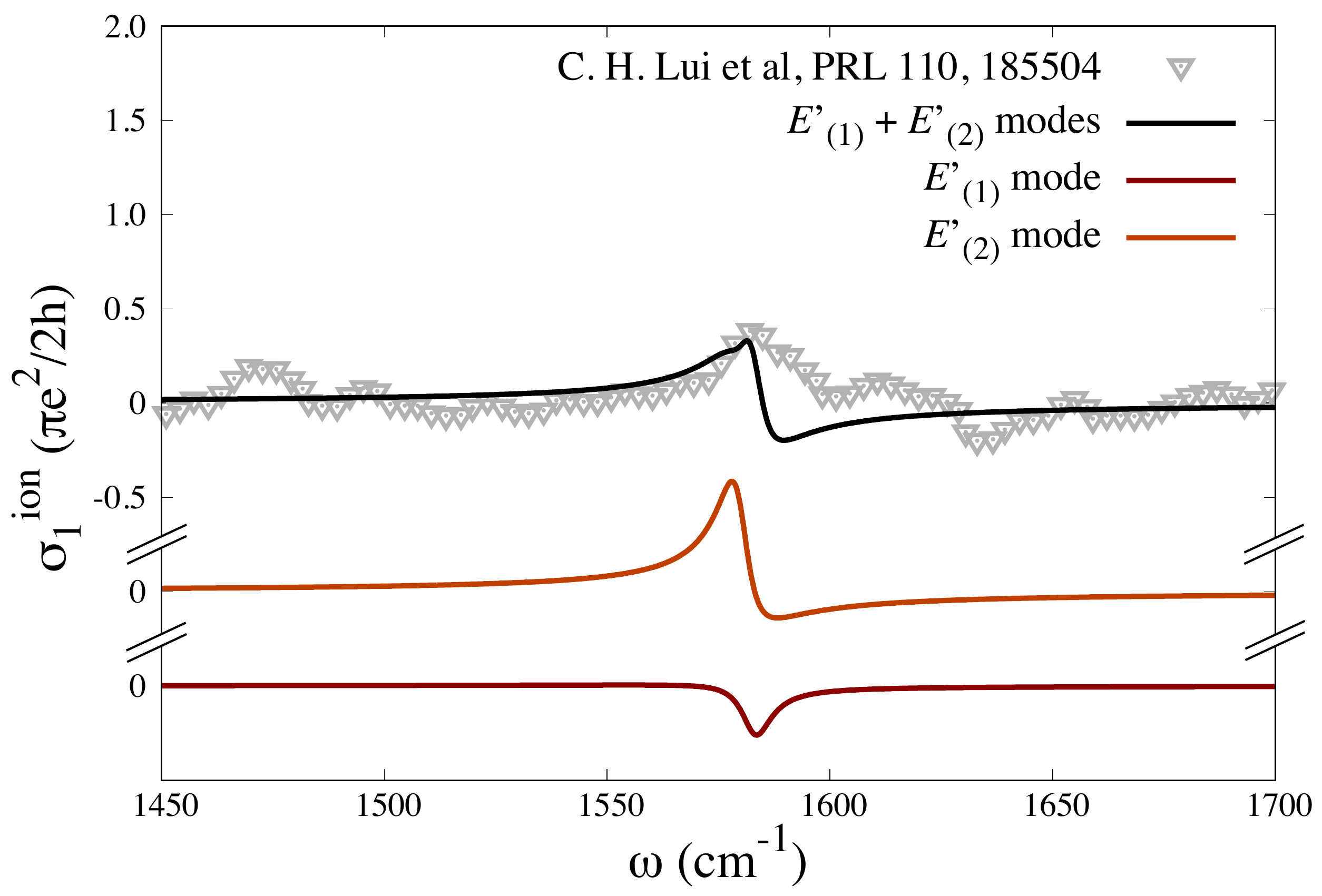}\label{8Fig-a}}\\
\subfloat[][Rhombohedral (ABC) stacking]{\includegraphics[width=8cm]{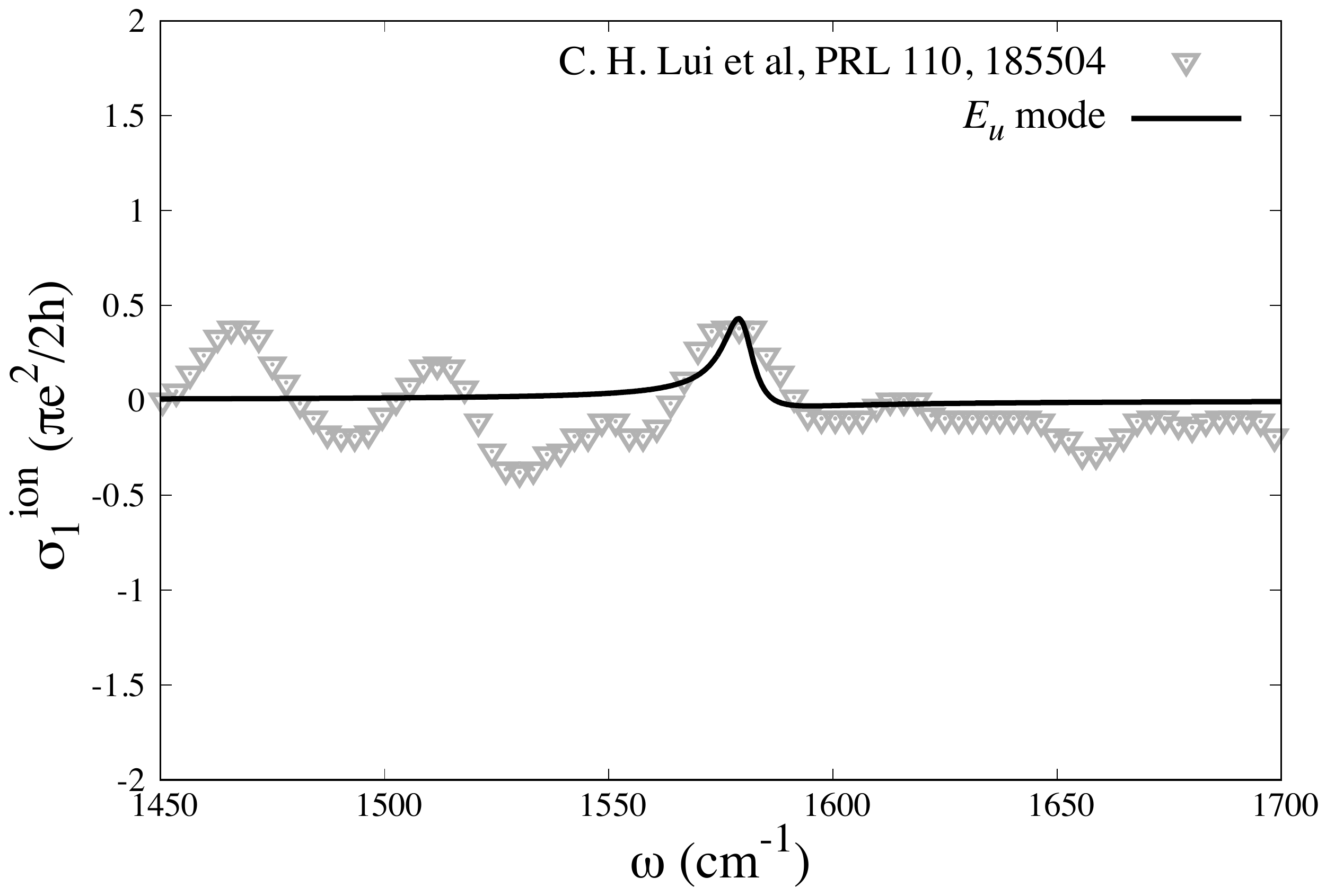}\label{8Fig-b}}
\caption{Comparison between simulated (continuous lines) and experimetal (dots) real part of the conductivity, from Ref. \cite{Lui2013}. In Fig. \ref{8Fig-a} we show, together with the total $\sigma_1^\mathrm{ion}$ (black), also the two contributions arising from the $E'_{(1)}$ (dark red) and $E'_{(2)}$ (orange) modes.}\label{8Fig}
\end{figure}

The calculated oscillator strengths and adiabatic/nonadiabatic frequencies for the two trilayers are reported in Table \ref{trilayer}. In Fig. \ref{8Fig} we compare our results with experimetal measurements from Ref. \cite{Lui2013}, showing the real part of the ionic conductivity $\bm{\sigma}^{\mathrm{ion}}(\omega)$ associated to the in-plane IR modes in units of $\pi e^2/(2h)$, linked to the ionic conductivity by the relation $\bm{\sigma}^{\mathrm{ion}}(\omega)=-i\omega\bm{\chi}^{\mathrm{ion}}(\omega)$. As $\bm{\chi}^{\mathrm{ion}}(\omega)$, also $\bm{\sigma}^{\mathrm{ion}}(\omega)$ is composed of a sum over IR optical modes, as shown in Ref. \cite{Bistoni2019}. In Fig. \ref{8Fig}, we upshifted all the peaks of $\approx 20$ cm$^{-1}$ -- that is 1\% of the NA phonon frequencies computed for the trilayers in Table \ref{trilayer} -- keeping the calculated \emph{ab initio} splittings of the IR $E'_{(1)}, E'_{(2)}$ modes shown in Fig. \ref{8Fig-a}. Notably, such IR modes have an almost opposite shape; when they are summed up $\sigma^{\mathrm{ion}}_\parallel(\omega\simeq0.2\,\mathrm{eV})\approx\sigma^{\mathrm{ion}}_{\parallel,E'_{(1)}}+\sigma^{\mathrm{ion}}_{\parallel,E'_{(2)}}$, the interference effect among them decreases the  overall IR intensity, which agrees better with the experimental data. Furthermore, we find that nonadiabatic renormalization effects, albeit small, may affect the splitting of the in-plane modes of trilayer graphene in a detectable way (e.g., by high-resolution Raman spectroscopy).   

We observe that the comparison of both the bilayer and the trilayer graphene with the experiments of Refs. \cite{Kuzmenko2009,Lui2013} is done for the zero-doping case. In the same articles, the authors studied also the case of finite doping $\varrho$, finding a substantial enhancement of the same IR peaks by increasing/decreasing the total number of electrons. We also mention an experimental study of the same IR resonance as a function of the number of graphene layers \cite{Li2012}, which however has been carried out with small unintentional doping levels. Although an analysis of the same peaks as a function of $\varrho$ would be interesting, it is beyond the scope of this article, but may be investigated in further studies.
\section{conclusions}  

In conclusion, we introduced an \emph{ab initio} scheme to describe the IR vibrational spectra of metallic crystalline solids in reflectivity measurements. We benchmarked our method by calculating the phonon signatures in the reflectivity spectra of graphite, finding good agreement between our results and available experimental data. We believe that our work will allow for a reliable first-principles description of reflectance spectra in metallic systems, in particular for those systems under extremely high-pressure conditions, as the new superconducting hydrides, where the IR vibrational spectroscopy  represents one of the few possible tools of investigation.

\subsection*{Acknowledgments}
We acknowledge Alexey Kuzmenko for sharing the in-plane data of reflectivity, and Michele Ortolani and Leonetta Baldassarre for a critical 
reading and useful suggestions. We also acknowledge financial support by the European Graphene Flagship Core 2 and Core 3 and the CINECA award under the ISCRA initiative (Grants HP10BSZ6LY and HP10BKBJMI) for the availability of high performance computing resources.

\appendix

\section{Expansion of reflectivity\label{approxR}}
The expansion of the reflectivity for $4\pi|\chi^{\text{I}}_{s,\alpha}|\ll| \epsilon^{\text{e}}_{s,\alpha}|$ is obtained by approximating
\begin{gather}
\sqrt{\epsilon_{s,\alpha}}=\sqrt{\epsilon^{\text{e}}_{s,\alpha}+4\pi\chi^{\text{I}}_{s,\alpha}}\sim \sqrt{\epsilon^{\text{e}}_{s,\alpha}}+\frac{1}{2} \frac{4\pi\chi^{\text{I}}_{s,\alpha}}{\sqrt{\epsilon^{\text{e}}_{s,\alpha}}}\\
\begin{split}
R_{s,\alpha}&=\left| \frac{\sqrt{\epsilon_{s,\alpha}}-1}{\sqrt{\epsilon_{s,\alpha}}+1}\right|^2\\
&\sim\left|\frac{\sqrt{\epsilon^{\text{e}}_{s,\alpha}}-1}{\sqrt{\epsilon^{\text{e}}_{s,\alpha}}+1}\right|^2\left|1+\frac{4\pi\chi^{\text{I}}_{s,\alpha}}{\sqrt{\epsilon^{\text{e}}_{s,\alpha}}(\epsilon^{\text{e}}_{s,\alpha}-1)}\right|^2.
\end{split}
\end{gather}
The expression given in Eq. (3) of the main text is recovered by setting $R^{\text{e}}_{s,\alpha}=\left|\frac{\sqrt{\epsilon^{\text{e}}_{s,\alpha}}-1}{\sqrt{\epsilon^{\text{e}}_{s,\alpha}}+1}\right|^2$ and neglecting again the quadratic term in the ratio $\frac{4\pi|\chi^{\text{I}}_{s,\alpha}|}{| \epsilon^{\text{e}}_{s,\alpha}|}$. For $\omega$ close to $\omega_s$, this gives the expression of Eq. (\ref{reflExp})
\begin{equation}
R_{s,\alpha}(\omega)=R^{\text{e}}_{s,\alpha}\left[1+2\,\textsf{Re}\,\bigg(\frac{4\pi\chi^{\text{I}}_{s,\alpha}(\omega)}{\sqrt{\epsilon^{\text{e}}_{s,\alpha}}\,(\epsilon^{\text{e}}_{s,\alpha}-1)}\bigg)\right]	\label{reflect}
\end{equation} 
  
\section{Implementation\label{implem}}
In this section we describe the technical details regarding the practical implementation of the effective charge tensor and electronic susceptibility within Quantum ESPRESSO \cite{Giannozzi2009,Giannozzi2017}. 

Neglecting the imaginary part of the induced charge density, the expression of the approximate frequency-dependent effective charge tensor can be written as \cite{Calandra2010}
\begin{equation}
\begin{split}
&{\bm{Z}}_\kappa^{\text{e}}(\omega_s)= -\frac{1}{N_\textbf{k}}\sum_{\textbf{k},nm}\sum_{\zeta=\pm}\frac{f_{\textbf{k},n}-f_{\textbf{k},m}}{  E_{\textbf{k},n}-E_{\textbf{k},m}+\zeta z_s} \\
&\times \big\langle u_{\textbf{k},m}\big|\,{\mathcal{V}}_\textsf{Re}^{\textbf{E}}(\omega_s)\big|u_{\textbf{k},n}\big\rangle\big\langle u_{\textbf{k},n}\big|\,{\mathcal{V}}_\textsf{Re}^{\textbf{u}_\kappa}(\omega_s) \big|u_{\textbf{k},m}\big\rangle\\
&+\frac{1}{\Omega}\int d^3 r \,d^3r'\,\textsf{Re}\,n^{\textbf{E}}(\textbf{r},\omega_s)\,K_{\text{Hxc}}(\textbf{r},\textbf{r}')\,\textsf{Re}\,n^{\textbf{u}_\kappa}(\textbf{r}',\omega_s)\label{effCharge2}
\end{split}
\end{equation}
where ${\mathcal{V}}_\textsf{Re}^{\bm{\xi}}(\omega)={V}^{\bm{\xi}}_{\text{KS}}\big[\textsf{Re}\,n^{\bm{\xi}}(\omega)\big]$, in which ${V}^{\bm{\xi}}_{\text{KS}}(\omega)$ is the Kohn-Sham potential, \emph{i.e.} the sum of the external and the Hxc potentials, perturbed with respect to a perturbation depending parametrically on $\bm{\xi}$. Note that ${\mathcal{V}}_\textsf{Re}^{\bm{\xi}}(\omega)={\mathcal{V}}_\textsf{Re}^{\bm{\xi}}(-\omega)$ is an hermitian operator, contrary to ${V}^{\bm{\xi}}_{\text{KS}}$ that instead satisfies ${V}^{\bm{\xi}}_{\text{KS}}(\omega)^\dagger={V}^{\bm{\xi}}_{\text{KS}}(-\omega)$. 

We rewrite the expression of the approximate effective charge tensor in the following way
\begin{equation}
\begin{split}
{\bm{Z}}_\kappa^{\text{e}}(\omega_s)&= -\frac{1}{N_\textbf{k}}\sum_{\textbf{k},n}\sum_{\zeta=\pm}\big\langle u_{\textbf{k},n}\big|\,{\mathcal{V}}_\textsf{Re}^{\textbf{E}}(\omega)\,{\mathcal{Q}}\big|u^{\textbf{u}_\kappa}_{\textbf{k},n,\zeta}(\omega)\big\rangle\\
&+\frac{1}{\Omega}\int d^3 r \,{\mathcal{V}}_\textsf{Re}^{\textbf{E}}(\textbf{r},\omega)\,\textsf{Re}\,n^{\textbf{u}_\kappa}(\textbf{r},\omega)
\end{split}
\end{equation}
where the sum over the band index $n$ can be restricted to the states with non-negligible occupations. The first-order expression of the wavefunction is
\begin{equation}
\begin{split}
{\mathcal{Q}}\big|u^{\textbf{u}_\kappa}_{\textbf{k},n,\zeta}(\omega)\big\rangle=\sum_m &\frac{f_{\textbf{k},n}-f_{\textbf{k},m}}{  E_{\textbf{k},n}-E_{\textbf{k},m}+\zeta z_s} \\
&\times \big\langle u_{\textbf{k},m}\big|\,{\mathcal{V}}_\textsf{Re}^{\textbf{u}_\kappa}(\omega)\big|u_{\textbf{k},n}\big\rangle\,|u_{\textbf{k},m}\rangle\label{PertWave}
\end{split}
\end{equation}
and the sum over $m$ can be restricted to states with negligible occupations, within which ${\mathcal{Q}}$ is the projector. We can divide this last space in two subspaces: a subspace whose energy bands are resonant with the frequency ({i.e.} for which there are two or more energies satisfying $E_{\textbf{k},m}-E_{\textbf{k},n}=\omega_s$) and a subspace whose energies are non-resonant with the electronic transitions (i.e. $E_{\textbf{k},m}-E_{\textbf{k},n}\neq\omega_s$ for every $m$ and $n$). In the former, a nonzero value of $\eta$ is necessary in order to avoid divergent integrand, whereas in the latter, it can be set $\eta=0$ since in this case, $ E_{\textbf{k},n}-E_{\textbf{k},m}\pm\omega\gg\pm\eta$ for every $n$, $m$ and \textbf{k}. The components of $\big|u^{\textbf{u}_\kappa}_{\textbf{k},n,\zeta}(\omega)\big\rangle$ which are resonant with $\omega_s$ can be treated by performing explicitly the sum in Eq. (\ref{PertWave}), while all of the other components (relative to the infinite-dimensional manifold of non-resonant states) can be computed by means of the Sternheimer equation \cite{Sternheimer1954} with the same strategy of Refs. \citep{Baroni2001,Baroni1987}. In this way, the first-order wavefunction can be obtained by
\begin{equation}
\begin{split}
{\mathcal{R}}\big|u^{\textbf{u}_\kappa}_{\textbf{k},n,\zeta}(\omega)\big\rangle=\sum_m& \frac{f_{\textbf{k},n}-f_{\textbf{k},m}}{  E_{\textbf{k},n}-E_{\textbf{k},m}+\zeta z_s}\,\\
&\times\big\langle u_{\textbf{k},m}\big|\,{\mathcal{V}}_\textsf{Re}^{\textbf{u}_\kappa}\big|u_{\textbf{k},n}\big\rangle\,|u_{\textbf{k},m}\rangle
\end{split}
\end{equation}
\begin{equation}
\begin{split}
(e^{-i\textbf{k}\cdot\textbf{r}}H_0e^{i\textbf{k}\cdot\textbf{r}}-E_{\textbf{k},n}&-\zeta\omega_s)\,{\mathcal{S}}\big|u^{\textbf{u}_\kappa}_{\textbf{k},n,\zeta}(\omega)\big\rangle\\
&=-f(E_{\textbf{k},n})\,{\mathcal{S}}\,{\mathcal{V}}_\textsf{Re}^{\textbf{u}_\kappa}|u_{\textbf{k},n}\rangle.
\end{split}
\end{equation}
in which ${\mathcal{Q}}={\mathcal{R}}+{\mathcal{S}}$; ${\mathcal{R}}$ is the projector onto resonant states and ${\mathcal{S}}$ is the projector onto the non-resonant subspace. Note that, with this scheme,  $\textsf{Im}\,{\bm{Z}}_\kappa^{\text{e}}$ embodies only contributions coming from resonant states.

The treatment for ${\bm{\epsilon}}^{\text{e}}(\omega_s)$ is similar to the one for ${\bm{Z}}_\kappa^{\text{e}}(\omega_s)$, with the only difference that, contrary to the effective charge tensor, it contains also a term corresponding to electronic intraband transitions, which at $\omega\rightarrow0$ manifests in the Drude peak. Therefore, we separate ${\bm{\epsilon}}^{\text{e}}(\omega_s)={\bm{\epsilon}}_{\text{inter}}^{\text{e}}(\omega_s)+{\bm{\epsilon}}_{\text{intra}}^{\text{e}}(\omega_s)$; ${\bm{\epsilon}}_{\text{inter}}^{\text{e}}(\omega_s)$ is calculated with the same methodology of ${\bm{Z}}_\kappa^{\text{e}}(\omega_s)$, and 
\begin{equation}
\begin{split}
{\bm{\epsilon}}_{\text{intra}}^{\text{e}}(\omega_s)=&\frac{8\pi(e\hbar)^2}{N_\textbf{k}\Omega}\sum_{\textbf{k},nm}\frac{d f(x)}{dx}\Big|_{x=E_{\textbf{k},n}}\\
&\times\frac{\langle u_{\textbf{k},n}|\textbf{v}_\textbf{k}|u_{\textbf{k},m}\rangle\,\langle u_{\textbf{k},m}|\textbf{v}_\textbf{k}|u_{\textbf{k},n}\rangle}{\omega_s^2-\eta_s^2+i2\omega_s\eta}
\end{split}
\end{equation}
is computed explicitly, summing over the occupied and the resonant bands.
\section{Brillouin zone sampling\label{kpoints}}
\begin{figure}[]
\centering
\includegraphics[width=7cm]{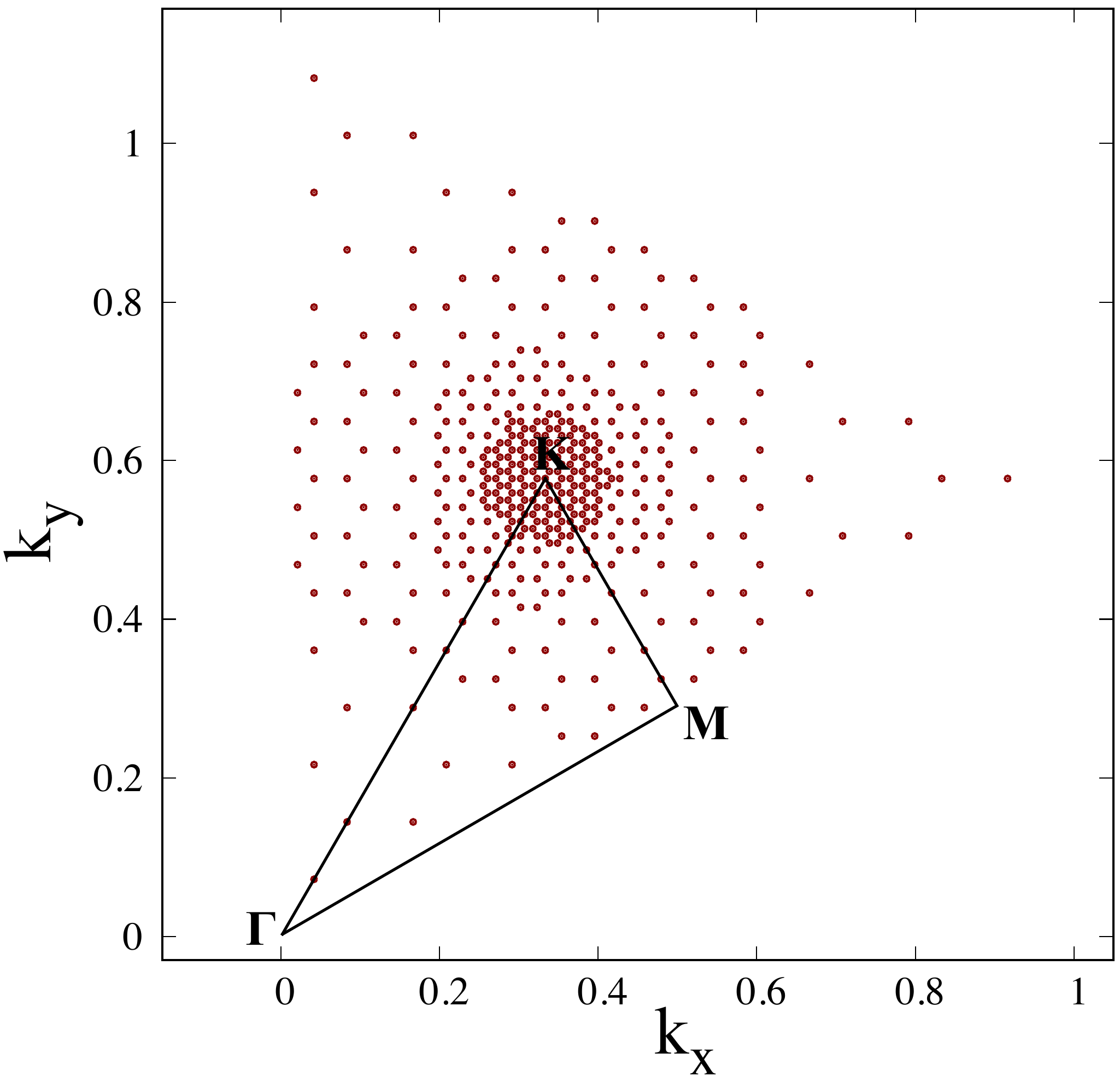}
\caption{Top-view of time-reversal symmetrized mesh used for the integration within the BZ. The highlighted region is the irreducible 2D wedge, which is the only part used for the \textbf{k}-point sums. This figure is obtained using $L=6$, $l=3$, $\mathcal{N}=8$ and, only for this figure, $p=2.1$.}\label{1Fig}
\end{figure}

Graphite is characterized by a density of states which decreases dramatically at the Fermi level. Because of its particular geometry of the band structure, it turns out that for frequency-dependent response function, like the effective charges or the electric susceptibility, most of the contributions to the \textbf{k}-point sum come from a small cylinder along the \textbf{K}-\textbf{H} line, in which the denominator of Eq. (\ref{effCharge2}) reaches its minimum value. Because of the very steep variation of the integrand around this region, an ultradense \textbf{k}-point grid is necessary to have a well-converged result; for this purpose, we employed a non-uniform \textbf{k}-points mesh in the $(k_x,k_y)$ plane of the Brillouin zone (BZ). Since the $(k_x,k_y)$ projection of the time-reversal-symmetrized BZ (TRS-BZ) of the hexagonal lattice is an equilateral triangle centered at \textbf{K}, we generate the grid in the following way: starting from one point at the centre of the triangle, we defined various levels of \textbf{k}-point densities; at the first level, from the point at the center of the BZ, we generated 3 points, each of them at the midpoint of the segment between the central point and one of the edges of the triangle. In this way, all the four points can be considered to be each of them at the center of a smaller triangle, whose area is $1/4$ of the area of the  $(k_x,k_y)$ TRS-BZ. The iteration of this procedure from all the four points produces $4^2$ points whose weight $w_\textbf{k}$ is $1/4^2$; this defines the second level. At the $\ell$-th level, the number of \textbf{k}-points is $4^\ell$ and $w_\textbf{k}=1/4^\ell$. In order to find the way in which the point should concentrate around the \textbf{K}-\textbf{H} line, \emph{i.e.} $\ell=\ell_{\textbf{k}}$, we required that, for a sum $\sum_\textbf{k}w_{\textbf{k}}\,I_{\textbf{k}}$, 
\begin{equation}
4^{-\ell_{\textbf{k}}} I_{\textbf{k}}=C
\end{equation}
where $C$ is a constant. Assuming that, far from \textbf{K}, $I_{\textbf{k}}\sim |{\textbf{k}-\textbf{K}}|^{-p}$, the density of \textbf{k}-points (which is fixed by $\ell$) at the distance $\mathcal{K}_\textbf{K}=|{\textbf{k}-\textbf{K}}|$ must be
\begin{equation}
\mathcal{K}_{\textbf{K}}(\ell)=C'\,4^{-\ell/p}.
\end{equation}
Defining $L$ as the max value of $\ell$, we fixed the constant $C'$ imposing $\mathcal{K}_\textbf{K}(L)={D}/\mathcal{N}$, where $\mathcal{N}$ is an integer and ${D}=4\pi/(3a)$ is the distance from $\textbf{K}$ to an edge of triangle; we choose $p=3$, according to the expression of $\textsf{Re}\,\bm{Z}_{\kappa}(\omega)$ of the model in Ref. \cite{Bistoni2019}; this choice is motivated by the fact that the linear response calculation is carried out with $\textsf{Re}\, n^{\bm{\xi}}$. As a result
\begin{equation}
\mathcal{K}_{\textbf{K}}(\ell;L,l,\mathcal{N})=\left(\frac{{D}}{\mathcal{N}}  \right)4^{(L-\ell)/3}
\end{equation}
where $\ell\in\{L,L-1,\dots,l+1,l\}$. The distance $\mathcal{K}_\textbf{K}$ depends parametrically on $L$, $l$ and $\mathcal{N}$, which must be seen as parameters over which convergence tests must be made. We reported in Fig. \ref{1Fig} an example of the non-uniform grid. 
\noindent

For the calculations we used $(L,l,\mathcal{N})=(9,4,25)$ and 90 $N_{k_z}$ for $\bm{Z}_\kappa^{\text{e}}(\omega)$, $(L,l,\mathcal{N})=(10,4,25)$ and 110 $N_{k_z}$ for $\bm{\epsilon}^{\text{e}}(\omega)$; these are equivalent, within the circle of radius $D/\mathcal{N}$, to a $724\times724\times90$ and a $1448\times1448\times110$ $\textbf{k}$-point grids, respectively.

\section{$\bm{\eta}$ dependence\label{depEta}}
\begin{figure}[h]
\centering
\captionsetup[subfigure]{labelformat=empty}
\subfloat[][\large{$E_{1u}$}]{\includegraphics[width=4cm]{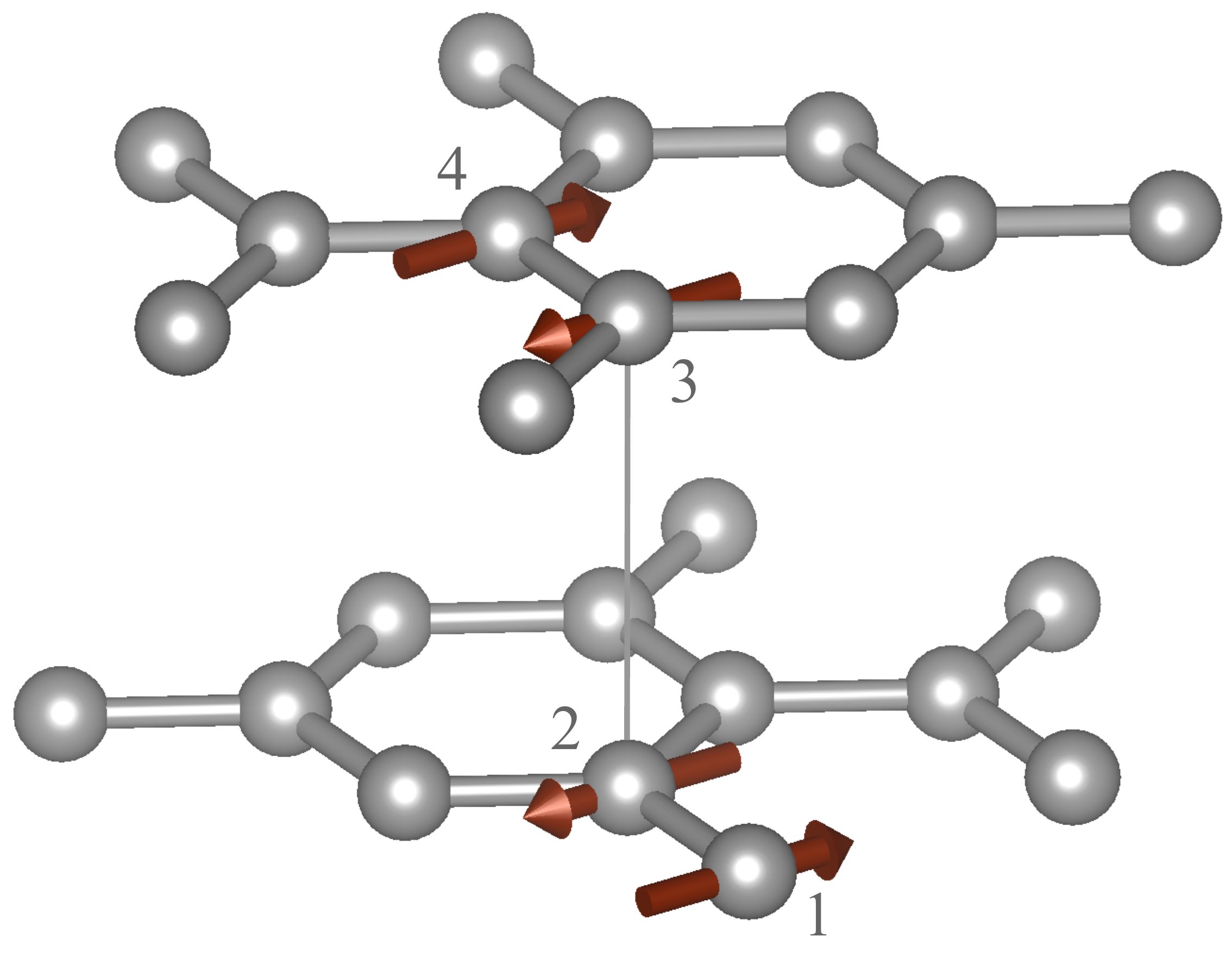}}
\subfloat[][\large{$A_{2u}$}]{\includegraphics[width=4cm]{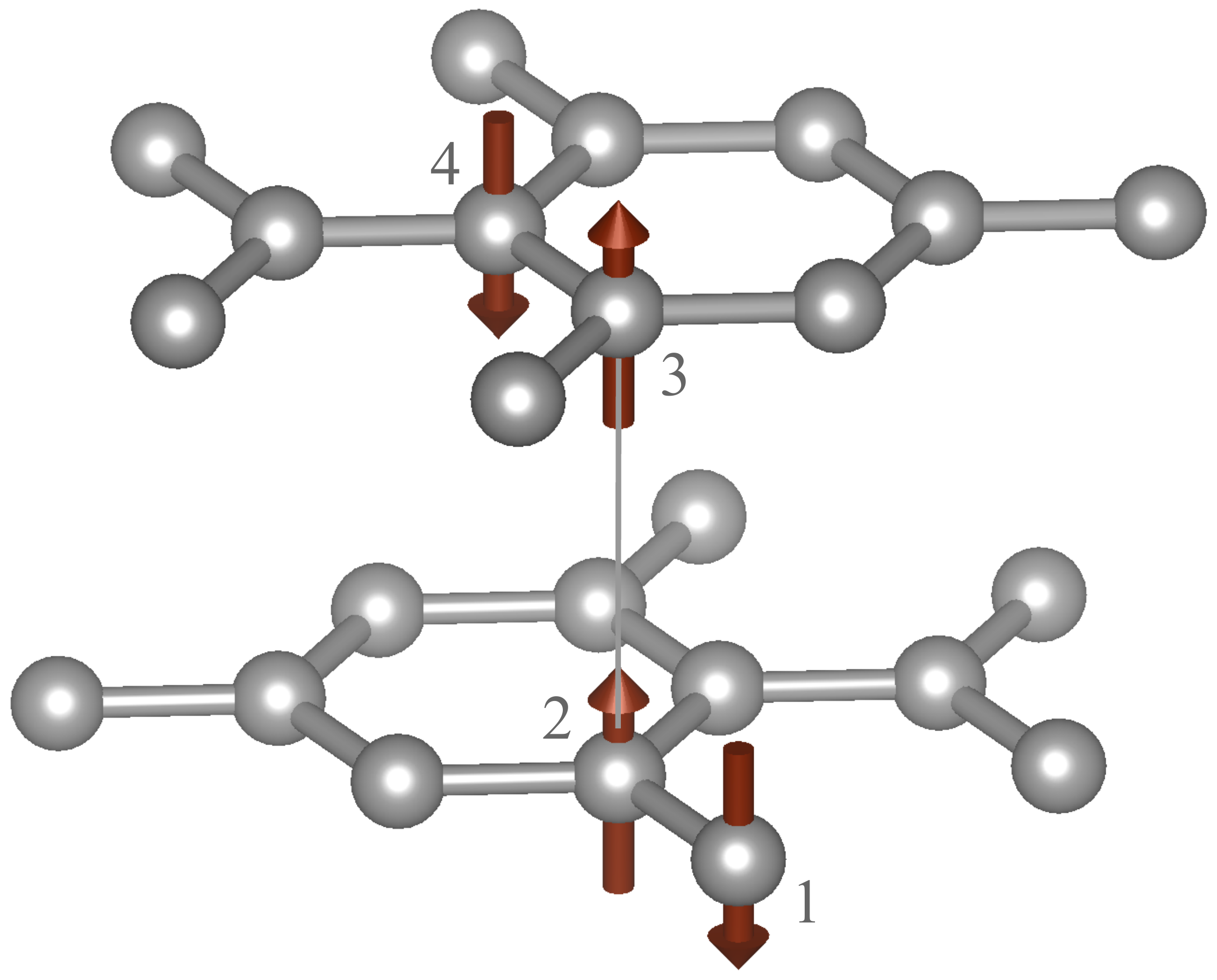}}
\caption{IR-active modes of graphite.}\label{5Fig}
\end{figure}

We considered the dependence of both $\bm{Z}_\kappa(\omega_s)$ and $\bm{\epsilon}^{\text{e}}(\omega_s)$ as a function of the parameter $\eta$. This parameter represents the inverse electronic lifetime, and it can be reasonably supposed to increase with the temperature, because of the increase with $T$ of both the electron-phonon and the electron-electron scattering processes.

\begin{table}[]
\centering
\caption{Dependence on $\eta$ of $\bm{Z}^{\text{e}}_\kappa(\omega_s)$ and $\bm{\epsilon}^{\text{e}}(\omega_s)$ for $s=E_{1u}$ (up) and $s=A_{2u}$ (down).}\label{tab1appendix}
\subfloat{
\begin{tabular}{*{7}{c}}
\midrule
\midrule
$\eta$ ($10^{-3}$ Ry)& $\textsf{Re}\,Z_{1,\parallel}$& $\textsf{Re}\,Z_{2,\parallel}$&$\textsf{Im}\,Z_{1,\parallel}$&$\textsf{Im}\,Z_{2,\parallel}$ &$\textsf{Re}\,\epsilon^{\text{e}}_\parallel$  &$\textsf{Im}\,\epsilon^{\text{e}}_\parallel$  \\
\midrule
 1  &  -0.268  &0.269 &-0.112&0.082&7.92 & 59.2  \\
  2 & -0.270 &0.273 &-0.105&0.075& 11.5 & 59.3 \\
 4& -0.267 &0.276 &-0.095&0.064&18.5&58.2\\
 6  &-0.263&0.276&-0.087&0.054& 25.1& 55.3\\
 8 &-0.258&0.276&-0.080&0.046& 30.4 & 51.1\\
\midrule
\midrule
\end{tabular}}\\
\subfloat{
\begin{tabular}{*{7}{c}}
\midrule
\midrule
$\eta$ ($10^{-3}$ Ry)& $\textsf{Re}\,Z_{1,\perp}$& $\textsf{Re}\,Z_{2,\perp}$&$\textsf{Im}\,Z_{1,\perp}$&$\textsf{Im}\,Z_{2,\perp}$ &$\textsf{Re}\,\epsilon^{\text{e}}_\perp$  &$\textsf{Im}\,\epsilon^{\text{e}}_\perp$  \\
\midrule
0.5 & -0.072&0.068&-0.00018 &0.00002 & 3.44 & 0.710 \\
 1  & -0.072&0.068&-0.00022 &0.00001 & 3.44 & 0.791 \\
 2  &  -0.072 &0.068&-0.00031 & 0.00000& 3.51 & 0.906 \\
 3  & -0.072&0.068&-0.00039 & -0.00001& 3.62 & 0.970 \\
 4  & -0.072&0.068&-0.00047 &-0.00002 & 3.73 & 0.980 \\
\midrule
\midrule
\end{tabular}}
\end{table}

We show in Fig. \ref{5Fig} the two IR-active mode of graphite. By symmetry, the effective charges of the two C atoms labeled by 2 and 3 are exactly equal, as well as the charges of the two 1 and 4 C atoms. The oscillator strenght $\textbf{d}_s(\omega_s)$ is obtained by taking the semidifference of $\bm{Z}_s(\omega_s)$, as detailed in the main text. In Table \ref{tab1appendix} we report the results of the calculations. The effective charge and the electronic dielectric tensors are both computed at $T=300$ K, with $a=2.46\;\text{\AA}$ and $c=6.70\;\text{\AA}$. All the other parameters (\emph{e.g.} cutoff, \textbf{k}-point mesh and so on) are the same as the previous calculations. It is found that $\textsf{Im}\,d_{E_{1u},\parallel}$ decreases with the increase of $\eta$; this fact suggests agreement with the experimental observation of the decrease of this quantity with the increase of $T$. 

 For the sake of completeness, we also report the numerical values for the effective charges at $T=150$ K evaluated with lattice constant $c=6.68$~\AA~ to account for the thermal compression, while keeping the in-plane lattice constant fixed at $a=2.46\;\text{\AA}$ \cite{Boettger1997}:
\begin{align*}
Z_{1,\parallel}(\eta=10^{-3} \,\text{Ry})&=-(0.267+i0.110)\\
Z_{2,\parallel}(\eta=10^{-3} \,\text{Ry})&=(0.270+i0.079)\\
Z_{1,\perp}(\eta=5\cdot10^{-4} \,\text{Ry})&=-(0.073+i0.00021)\\
Z_{2,\perp}(\eta=5\cdot10^{-4} \,\text{Ry})&=(0.071+i0.00002)
\end{align*}
Overall, all considered quantities do not show significant dependences on thermal expansion, whereas they depend more markedly on the damping parameter $\eta$.

\section{Dielectric tensor within IR range}
In this section we show the results of our calculation of the frequency-dependent dielectric tensor in a wider range of IR spectrum. Using $\eta_{E_{1u}}=110$ cm$^{-1}$ and the same input files as for the previous computations, we calculated the electronic dielectric tensor from 1100 to 1900 cm$^{-1}$. We then evaluated $R^{\mathrm{e}}_{E_{1u},\parallel}$ using the computed $\epsilon_{\parallel}^{\text{e}}(\omega_{E_{1u}})$. 

\begin{figure}[h]
\centering
\includegraphics[width=8cm]{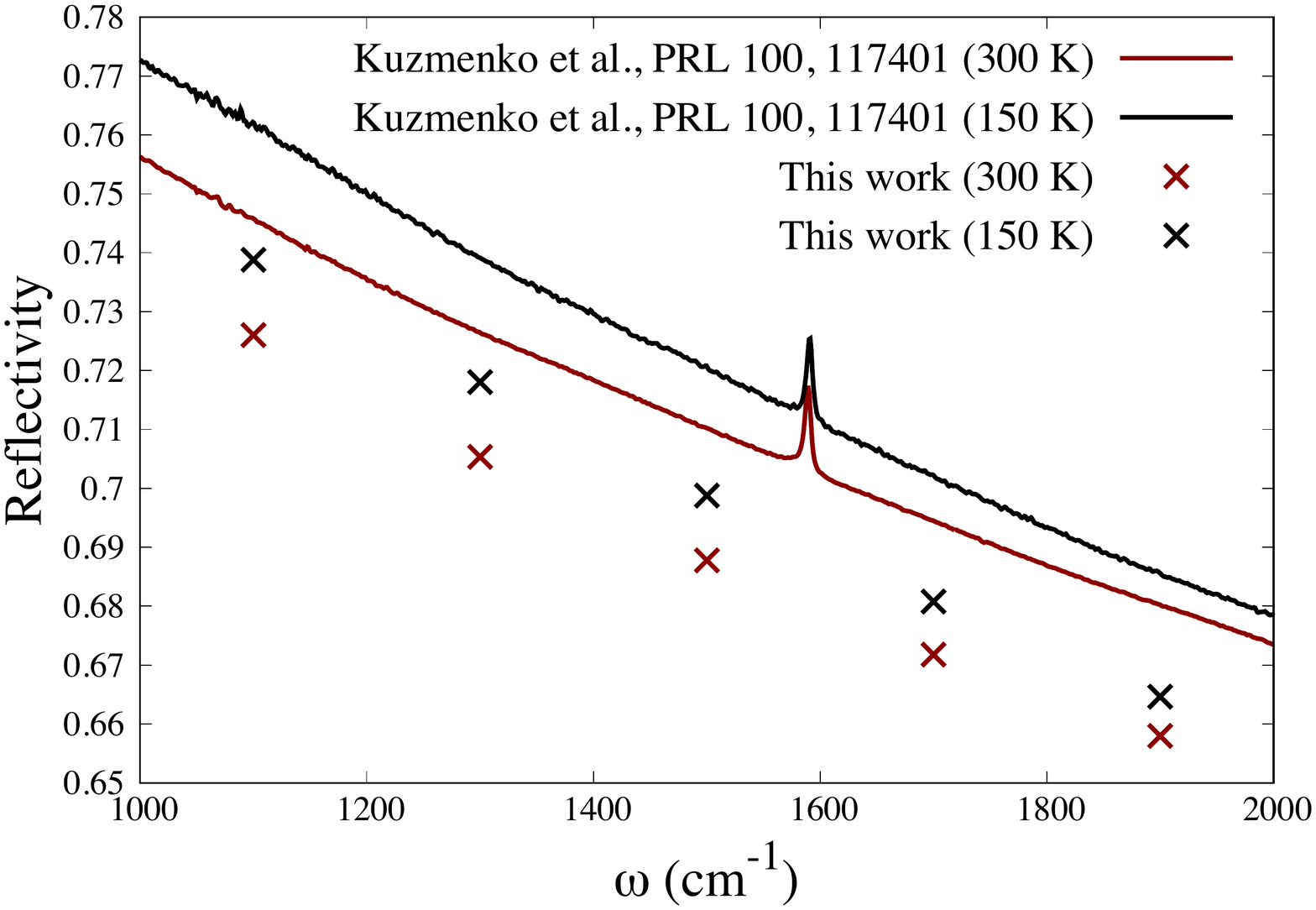}
\caption{Frequency dependence of $R^{\text{e}}_{s,\alpha}$ in Eq.(\ref{reflect}).}\label{electrRefl}
\end{figure}

We show in Fig. \ref{electrRefl} the results of our calculations, compared to the experimental measurements of reflectivity of Ref. \cite{Kuzmenko2008}. The calculated and the experimental $R^{\mathrm{e}}_{E_{1u},\parallel}$, besides having the same $\omega$-dependence, also display a very similar dependence on temperature, that becomes less important at larger frequencies.
\bibliography{1_bibliography}
\end{document}